\documentclass[12pt]{article}
\usepackage{color}
\usepackage{stmaryrd}
\usepackage{bbm}
\usepackage{mathrsfs}
\usepackage{amsfonts}
\usepackage{amssymb}
\usepackage{amsmath}
\usepackage{amsthm}
\usepackage{cases}
\usepackage{indentfirst}
\usepackage{graphicx}
\usepackage{caption}
\usepackage{float}
\usepackage{epstopdf}
\usepackage{algorithm}
\usepackage{algorithmic}
\usepackage[sort&compress,numbers]{natbib}
\usepackage{amsmath,amssymb,mathrsfs,amscd,graphicx,float,subfigure}

\newtheorem{remark}{Remark}
\allowdisplaybreaks[4]

\def\be{\begin{eqnarray}} \def\ee{\end{eqnarray}} 
  \def\({\left(} \def\){\right)}
\def\bc{\begin{center}} 
\def\ec{\end{center}}  
\def\bey{\begin{eqnarray*}}\def\eey{\end{eqnarray*}}

\begin{document}

\title{ {\bf  Meshy soliton structures for  (2+1)-dimensional integrable systems and interactions}
\footnotetext{* Corresponding Author}}
\author{Shoufeng Shen,  \qquad Guofang Wang, \qquad Yongyang Jin $^{*}$\\
}
\date{}
\maketitle
\begin{center}
\begin{minipage}{135mm}
{\small  Department of Applied Mathematics, Zhejiang University of
Technology, Hangzhou 310023, China}\\
\end{minipage}
\end{center}


\noindent{\bf Abstract:}

In this letter,  we construct new meshy soliton structures by using two concrete (2+1)-dimensional integrable systems.
The explicit expressions based on corresponding Cole-Hopf type transformations are obtained.
Constraint equation $f_t+\sum_{j=1}^N h_j(y)f_{jx}=0$ shows that these meshy soliton structures  can be linear or parabolic.
Interaction between meshy soliton structure and Lump structure are also revealed.

\vspace {3mm}


\vspace {3mm}

\noindent{\bf PACS numbers:}

\vspace {2mm}

\section{Introduction}

The concept of soliton introduced  by Zabusky and Kruskal,  is originally used to describe  the collision of two localized KdV solitary waves in FPU problem. Nowadays, solitons are receiving much attention in many natural sciences such as biology, chemistry, mathematics, communication, and
especially in almost all branches of physics like condense matter physics, quantum field theory, plasma physics, fluid mechanics and nonlinear optics. Usually, one considers that the solitons are the basic excitations of integrable
systems while the chaos and fractals are the basic behaviors
of non-integrable systems. Thus, one of the most important research
fields in soliton theory is to seek for exact solutions of integrable systems and use these solutions to simulate various natural science phenomena \cite{iss, serh, llka, zhh, mac, wtl}.

 Recently,  many new (2+1)-dimensional soliton structures such as dromions, lumps, ring solitons, breathers, instantons, compactons have been obtained. These studies are restricted in the single-valued situations. For more complicated cases, multivalued-functions also
have been used to construct folded solitary waves and foldons. For example, the dromion structures  which decay exponentially in all directions, can be obtained by two non-perpendicular line solutions for the Kadomtsev-Petviashvilli (KP) equation, while for the Davey-Stewartson (DS) and Nizhnik-Novikov-Veselov (NNV) equations,  the dromion solutions are constructed by two perpendicular line solutions.

\begin{figure}[h]
 \centering
\includegraphics[width=7cm,height=5cm]{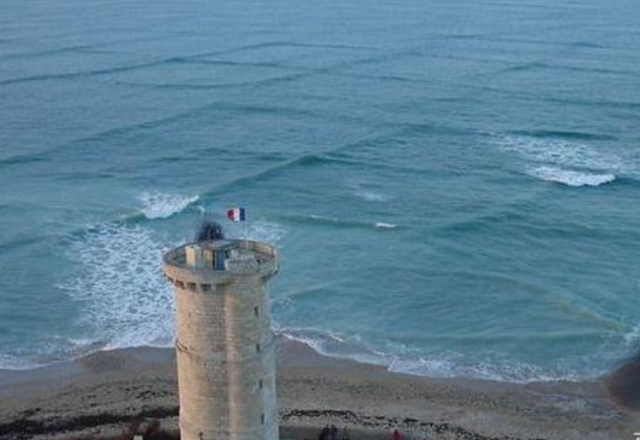}
\captionsetup{font={footnotesize}}
\caption{\footnotesize {Meshy  wave on a sea surface in France.} }
\label{f0-png}
\end{figure}

In this letter,  we construct new soliton structures, which are called the meshy soliton structures by using two concrete (2+1)-dimensional integrable systems. It should be emphasized that the meshy soliton structure was first named and studied by Prof. Yong Chen. The obtained results may be able to simulate and explain the  water wave phenomenon in Internet figure \ref{f0-png}, which has happened on a sea surface in France.

\section{Meshy soliton structures  and interactions}

We focus on the following coupled nonlinear
system \cite{wlt,waz},
\begin{equation}\label{a1}
\left\{\begin{array}{l}
u_t-2uu_x-v_{xx}=0,\vspace{2mm}\\
v_{yt}-u_{xxy}-2uv_{xy}-2u_xv_y=0,
  \end{array}\right.
\end{equation}
which deserves the name as a coupled Burgers system because
by setting $u = v$ and $x = y$, it is reduced to the standard
Burgers equation.
Substituting
\begin{equation}\label{a2}
\left\{\begin{array}{l}
u=\frac{f_x}{f}+h(y),\vspace{2mm}\\
v=u+\alpha,
  \end{array}\right.
\end{equation}
into the above system, we have the following linear equation,
\begin{equation}\label{a3}
f_t-2h(y)f_x-f_{xx}=0.
\end{equation}
Here $f\equiv f(x,y,t)$ and $\alpha$ is an arbitrary constant.
Obviously, Eq.\eqref{a2} is a Cole-Hopf  type transformation which transform the original coupled Burgers system \eqref{a1}
to the heat conduction type equation \eqref{a3}.
Moreover, another integrable coupled system of the modified KdV equation and the potential
Boiti-Leon-Manna-Pempinelli equation (mKdV-pBMLP) is given in Refs.\cite{wlt,waz} as
\begin{equation}\label{b1}
\left\{\begin{array}{l}
u_t + 3u^2u_x + u_{xxx} + \frac{3}{2}(uv_x)_x + \frac{3}{2}\partial^{-1}_y (uv_y)_{xx}=0,\vspace{2mm}\\
v_{yt} + \frac{3}{2}(v_xv_y)_x + v_{xxxy} + 3(u^2v_y)_x + \frac{3}{2}(uu_y)_{xx} +\frac{3}{2}(uu_{xy})_x=0.
  \end{array}\right.
\end{equation}
By means of Eq.\eqref{a2}, we have the following linear equation
\begin{equation}\label{b2}
f_t+3h^2(y)f_x+3h(y)f_{xx}+f_{xxx}=0.
\end{equation}
Thus we can construct new meshy soliton structures for the physical quantity
\begin{equation}\label{b3}
U=\gamma({\rm ln} f)_{xy}=\gamma\frac{f_{xy}f-f_xf_y}{f^2},
\end{equation}
via suitable selections of function $f$ in Eq.\eqref{a3} and Eq.\eqref{b2}.
Here $\gamma$ is a constant, just for the sake of computer simulation and drawing.
In this letter, we can take the value of $\gamma$ to be $-1$.

If we set $f=1+\sum_{j=1}^{N}c_j e^{a_j x+b_j y+a_j^2 t+d_j}$ in Eq.\eqref{a3} with $h(y)=0$, where $a_j, ~b_j, ~ c_j, ~d_j$ are appropriate constants, then we have  dromion structure or multiple solitoff structure. Here we call a half-straight line soliton structure as a solitoff. These conclusions have been presented in many literatures.
Thanks to the arbitrariness of the constants $b_j$, we can construct new Meshy soliton structure displayed in Figures 2-7 by setting
\begin{eqnarray}\label{d1}
&& f_1=1+e^{-4x+y+16t}+e^{4x-y+16t-30}+e^{-12x+3y+144t-50}+e^{x-4y+t}+e^{-3x-3y+9t}\nonumber\\
&& \qquad\quad +e^{5x-5y+25t-30}+e^{-11x-y+121t-50}+e^{-x+4y+t-30}+e^{-5x+5y+25t-30}\nonumber\\
&& \qquad\quad +e^{3x+3y+9t-60}+e^{-13x+7y+169t-80}+e^{3x-12y+9t-50}+e^{-x-11y+t-50}\nonumber\\
&& \qquad\quad +e^{7x-13y+49t-80}+e^{-9x-9y+81t-100}.
\end{eqnarray}
From Figures 2-4 to Figures 5-7, we can see that the shape of meshy soliton structure is most regular as the time variable goes to $0$ and becoming a parallelogram type soliton structure. Theoretically, we give the interpretation of water wave simulation in Figure 1.

\begin{figure}[H]
\begin{minipage}[t]{0.3\linewidth}
\centering
\includegraphics[width=7cm,height=5cm]{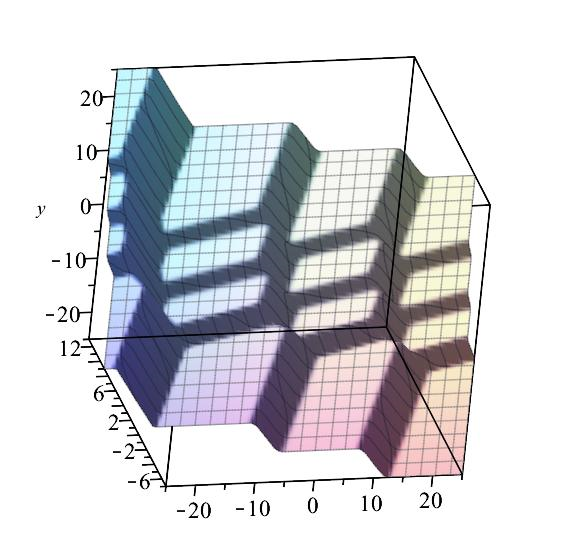}
\captionsetup{font={footnotesize}}
\caption{\footnotesize  Meshy kink-soliton structure of the field function $-u$  in Eq.\eqref{a2} by using $f_1$ at $t=-1$.}
\end{minipage}\qquad
\begin{minipage}[t]{0.3\linewidth}
\centering
\includegraphics[width=7cm,height=5cm]{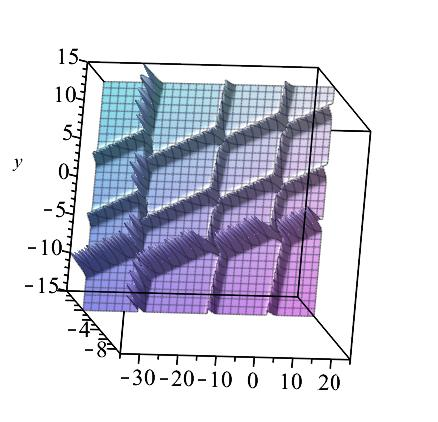}
\captionsetup{font={footnotesize}}
\caption{\footnotesize Meshy soliton structure  of $U$ by using  $f_1$ at  $t=-1$.}
\end{minipage}\qquad
\begin{minipage}[t]{0.3\linewidth}
\centering
\includegraphics[width=6cm,height=4.5cm]{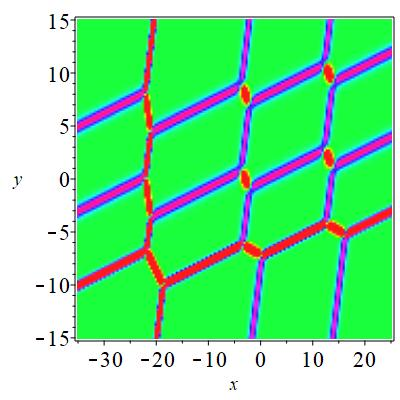}
\captionsetup{font={footnotesize}}
\caption{\footnotesize Density plot of  $U$ by using  $f_1$ at $t=-1$.}
\end{minipage}
\end{figure}

\begin{figure}[H]
\begin{minipage}[t]{0.3\linewidth}
\centering
\includegraphics[width=7cm,height=5cm]{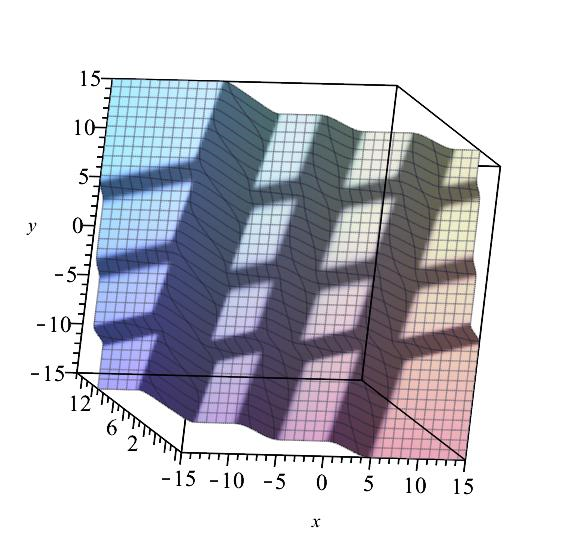}
\captionsetup{font={footnotesize}}
\caption{\footnotesize Meshy kink-soliton structure of the field function $-u$ in Eq.\eqref{a2} by using $f_1$ at $t=0$}
\end{minipage}\qquad
\begin{minipage}[t]{0.3\linewidth}
\centering
\includegraphics[width=7cm,height=5cm]{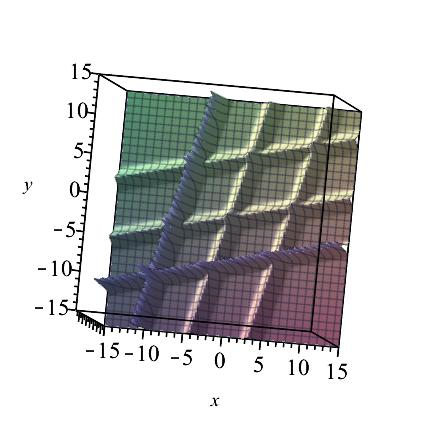}
\captionsetup{font={footnotesize}}
\caption{\footnotesize Meshy soliton structure  of  $U$ by using  $f_1$ at  $t=0$.}
\end{minipage}\qquad
\begin{minipage}[t]{0.3\linewidth}
\centering
\includegraphics[width=6cm,height=4.5cm]{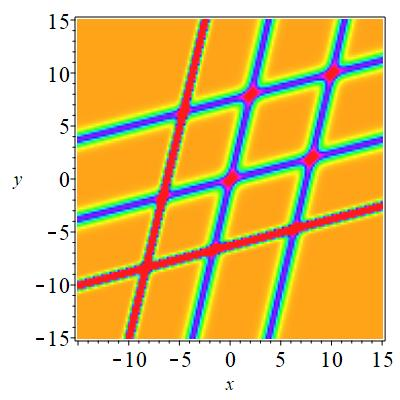}
\captionsetup{font={footnotesize}}
\caption{\footnotesize Density plot of  $U$ by using  $f_1$ at $t=0$.}
\end{minipage}
\end{figure}

To consider furthermore,  we write another kind of meshy soliton structure by using the following functional expression \eqref{e1}, which  is composed of linear solitons and parabolic solitons.
\begin{eqnarray}\label{e1}
&& f_2=1+e^{x-4y+t}+e^{-x+4y+t-30}+e^{3x-12y+9t-50}+e^{-4x+y^2+16t}+e^{-3x-4y+y^2+9t}\nonumber\\
&& \qquad\quad +e^{-5x+4y+y^2+25t-30}+e^{-x-12y+y^2+t-50}+e^{4x-y^2+16t-30}+e^{5x-4y-y^2+25t-30}\nonumber\\
&& \qquad\quad +e^{3x+4y-y^2+9t-60}+e^{7x-12y-y^2+49t-80}+e^{-12x+3y^2+144t-50}+e^{-11x-4y+3y^2+121t-50}\nonumber\\
&& \qquad\quad +e^{-13x+4y+3y^2+169t-80}+e^{-9x-12y+3y^2+81t-100}.
\end{eqnarray}
For the sake of readability and simplicity, we present only two figures  to illustrate this soliton structure (see Figures 8-9).

\begin{figure}[H]
\qquad\qquad\qquad\begin{minipage}[t]{0.3\linewidth}
\centering
\includegraphics[width=7cm,height=5cm]{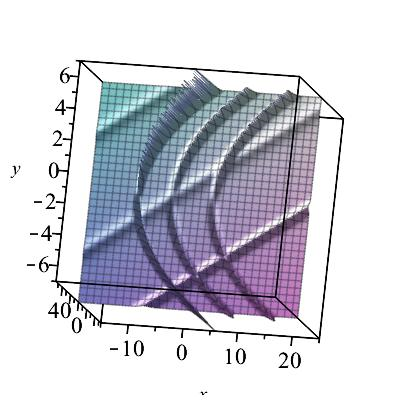}
\captionsetup{font={footnotesize}}
\caption{\footnotesize Meshy soliton structure  of $U$ by using  $f_2$ at  $t=0$}
\end{minipage}\qquad
\begin{minipage}[t]{0.45\linewidth}
\centering
\includegraphics[width=6cm,height=4.5cm]{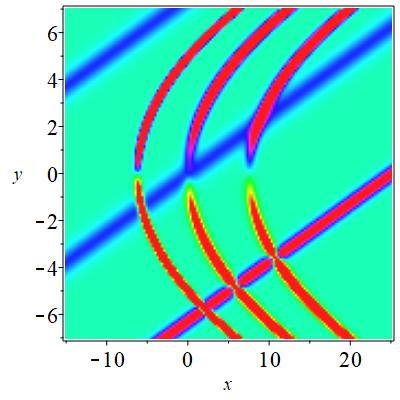}
\captionsetup{font={footnotesize}}
\caption{\footnotesize Density plot of   $U$ by using  $f_2$ at $t=0$}
\end{minipage}
\end{figure}

 Finally, we construct the interaction  behavior between meshy soliton structure and Lump structure by using the following functional expression \eqref{g1},
\begin{eqnarray}\label{g1}
&& f_3=f_1+x^2+2t+y^2-\frac{39}{20}.
\end{eqnarray}
Figures 9-10 show the  interaction  behavior at $t=0$, which has parallelogram type soliton-Lump structure.
For $f_4=f_2+x^2+2t+y^2-\frac{39}{20}$, we have an almost similar result.

\begin{figure}[H]
\qquad\qquad\qquad\begin{minipage}[t]{0.3\linewidth}
\centering
\includegraphics[width=7cm,height=5cm]{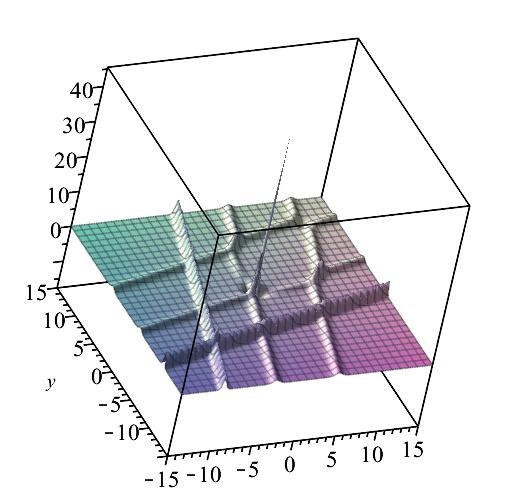}
\captionsetup{font={footnotesize}}
\caption{\footnotesize Interaction  behavior between meshy soliton structure and Lump structure of $U$ by using  $f_3$ at $t=0$. }
\end{minipage}\qquad
\begin{minipage}[t]{0.45\linewidth}
\centering
\includegraphics[width=6cm,height=4.5cm]{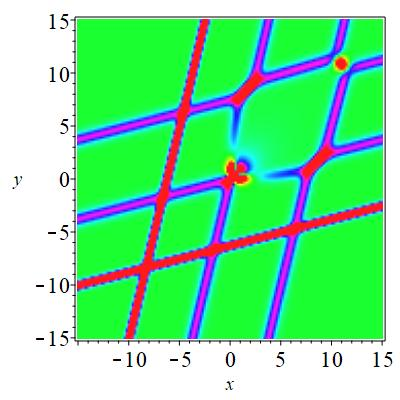}
\captionsetup{font={footnotesize}}
\caption{\footnotesize Density plot of   $U$ by using  $f_3$ at $t=0$.}
\end{minipage}
\end{figure}

\begin{remark}
The choice of function $f\equiv f(x,y,t)$ depends on the constraint equations \eqref{a3} or \eqref{b2}.
In fact, starting from the following generalized constraint  equation
\begin{eqnarray}\label{m1}
&& f_t+\sum_{j=1}^N h_j(y)f_{jx}=0
\end{eqnarray}
and corresponding Cole-Hopf type transformation, we can construct some new integrable systems from the inverse problem point of view.
We can also consider differential-difference equations and  nonlocal constraint equations such as $f_t+h(y)f_x(-x)+f_{xxx}=0$.
\end{remark}

\begin{remark}
We can also study meshy peakon structure and meshy loop-soliton structure.
\end{remark}

\section{Conclusions}

We considered  the (2+1)-dimensional coupled Burgers and mKdV-pBLMP system, which are reduced to linear constraint equations \eqref{a3} and \eqref{b2}. Although the construction of (2+1)-dimensional soliton structures is more difficult than that of (1+1)-dimensional soliton structures \cite{sjz, qhmsw},
new meshy soliton structures represented by Figures 2-8 for the physical quantity $U$ are obtained, which can be linear or parabolic. These results may be used to simulate the  water wave phenomenon in Figure \ref{f0-png}, which has happened on a sea surface in France.
In Figures 9-10, interaction between meshy soliton structure and Lump structure are also revealed.

\section*{Acknowledgments}

The authors thank Prof. Yong Chen of East China Normal University for helpful discussions.
The work was supported
by the National Natural Science Foundation of China (11771395).

\end{document}